\documentclass[preprint,showpacs]{revtex4}
\usepackage{graphicx}
\begin{document}
\title{Numerical study of roughness distributions in
nonlinear models of interface growth}
\author{F. D. A. Aar\~ao Reis}
\address{
Instituto de F\'\i sica, Universidade Federal Fluminense,\\
Avenida Litor\^anea s/n, 24210-340 Niter\'oi RJ, Brazil}
\date{\today}

\begin{abstract}
We analyze the shapes of roughness distributions of discrete
models in the Kardar, Parisi and Zhang (KPZ) and in the 
Villain, Lai and Das Sarma (VLDS) classes of interface growth,
in one and two dimensions. Three KPZ models in $d=2$ confirm
the expected scaling of the distribution and show a
stretched exponential tail approximately as $\exp{\left( -x^{0.8}\right)}$,
with a significant asymmetry near the maximum. Conserved restricted
solid-on-solid models belonging to the VLDS class were
simulated in $d=1$ and $d=2$. The tail in $d=1$ has the form
$\exp{\left( -x^2\right)}$ and, in $d=2$, has a simple exponential decay, but
is quantitatively different from the distribution of the linear
fourth-order (Mullins-Herring) theory. It is not possible to fit any of the
above distributions to those of $1/f^\alpha$ noise interfaces, in contrast with
recently studied models with depinning transitions.
\end{abstract} \date{\today}

\maketitle
\narrowtext

\section{Introduction}
\label{intro}

The simplest quantitative characteristic of an interface is its width, or
roughness, $w$, defined as the rms fluctuation of the height around its
average position. The scaling of the average width of different
realizations with the length $L$ and the time $t$ is useful
for analyzing interfaces formed in several processes~\cite{barabasi,krug}.
Another potentially useful characteristic is the full roughness distribution
at the steady state, particularly in cases where reliable estimates of scaling
exponents are not available.

The distributions for linear growth models, such as the Edwards-Wilkinson
(EW) and the Mullins-Herring (MH)
equations, were obtained exactly in $d=1$ and can be computed with high
degree of accuracy in $d=2$~\cite{foltin,plischke,racz}. The generating
functions for Gaussian interfaces with power spectrum of the type
$1/f^\alpha$ were computed by the same
methods~\cite{antal1,antal}. However, many real interfaces are described by
nonlinear equations at a coarse-grained level, and their
scaling exponents and roughness distributions are exactly known only in
some particular cases. Consequently, their accurate numerical calculation may
provide a basis for comparison with results from experiments or model
systems. In this Brief Report, we will study numerically the
distributions of models governed by second and fourth order nonlinear growth
equations, in $d=1$ and $d=2$.

The first model is that
of Kardar, Parisi and Zhang (KPZ)~\cite{kpz}, in which the local height $h$ at
position $\vec{x}$ and time $t$ evolves as
\begin{equation}
{{\partial h}\over{\partial t}} = \nu_2{\nabla}^2 h + \lambda_2
{\left( \nabla h\right) }^2 + \eta (\vec{x},t) .
\label{kpz}
\end{equation}
Here, $\nu_2$ and $\lambda_2$ are constants and $\eta$ is a Gaussian
noise with zero mean and variance $\langle
\eta\left(\vec{x},t\right) \eta (\vec{x'},t') \rangle = D\delta^d
(\vec{x}-\vec{x'}) \delta \left( t-t' \right)$, where $d$ is the
dimension of the substrate. In the case $\lambda_2=0$, we obtain the
EW equation. 

Another important nonlinear model is that of Villain~\cite{villain} and Lai and
Das Sarma~\cite{laidassarma}, the so-called VLDS equation (or nonlinear MBE
equation)
\begin{equation}
{{\partial h}\over{\partial t}} = \nu_4{\nabla}^4 h +
\lambda_{4} {\nabla}^2 {\left( \nabla h\right) }^2 + \eta (\vec{x},t) ,
\label{vlds}
\end{equation}
where $\nu_4$ and $\lambda_{4}$ are constants. The linear version of Eq.
(\ref{vlds}) ($\lambda_4 =0$) is the MH
equation~\cite{barabasi,plischke,racz,mullins,herring}.

Previous studies~\cite{foltin,racz} of the roughness distributions
of discrete models in these classes suggested that they have a universal
shape, with the form
\begin{equation}
P_L\left( w_2\right) = {1\over \langle w_2\rangle}
\Phi\left( {w_2\over{\langle w_2\rangle}}\right) ,
\label{scaling2}
\end{equation}
where $P_L\left( w_2\right) dw_2$ is the probability that the width
of a given configuration lies in the range $\left[ w_2, w_2+dw_2 \right]$.
Alternatively, one may use the scaling form
\begin{equation}
P_L\left( w_2\right) =
{1\over \sigma} \Psi\left( {{w_2-\left< w_2\right>}\over\sigma}\right) ,
\label{scaling1}
\end{equation}
where $\sigma \equiv \sqrt{ \left< {w_2}^2 \right> - {\left< w_2\right>}^2 }$
is the rms deviation of the squared width.

For KPZ in $d=1$, these scaling relations are analytically predicted and
confirmed by numerical data~\cite{foltin}. In $d=2$, simulations of discrete
models showed agreement with the scaling relation
(\ref{scaling2})~\cite{racz,marinari2}, but the shape of the
scaling function was not analyzed in detail.
Here, we will calculate accurate roughness distributions for three
discrete models in $d=2$: the etching model of Mello et
al~\cite{mello}, the restricted solid-on-solid (RSOS) model of Kim and
Kosterlitz~\cite{kk} and a generalized RSOS model with maximum neighboring
heights difference ${\Delta H}_{max}=2$~\cite{kk,alanissila}, hereafter called
RSOS2. Their analysis is presented in Sec. II.

Roughness distributions of the VLDS class were previously
obtained by R\'acz and Plischke~\cite{racz}, who simulated a model with
Arrhenius dynamics in $d=2$. In order to analyze the
shapes of the VLDS distributions, both in $d=1$ and $d=2$, we will
simulate conserved solid-on-solid (CRSOS) models which
show moderate finite-size effects when compared to other discrete VLDS
models. Their original versions were proposed by Kim et al~\cite{crsos1} and
were rigorously proved to belong to the VLDS class~\cite{huang,park}. Their
analysis is presented in Sec. III.

In Sec. IV we summarize our results and conclusions.

\section{Roughness distributions of KPZ models in $2+1$ dimensions}
\label{seckpz}

In Fig. 1 we show the distributions of KPZ models scaled according to Eq.
(\ref{scaling1}), with $x\equiv \left( w_2-\left< w_2\right>\right)/ \sigma$.
Results for other system sizes confirm the expected scaling of Fig. 1. The
number of configurations in which $w_2$ was measured was near ${10}^8$ for
each model and periodic boundary conditions were adopted. Eq. (\ref{scaling2})
and (\ref{scaling1}) are equivalent, but the latter was used because
it provides much better data collapse for the KPZ models.

\begin{figure}[h]
\includegraphics{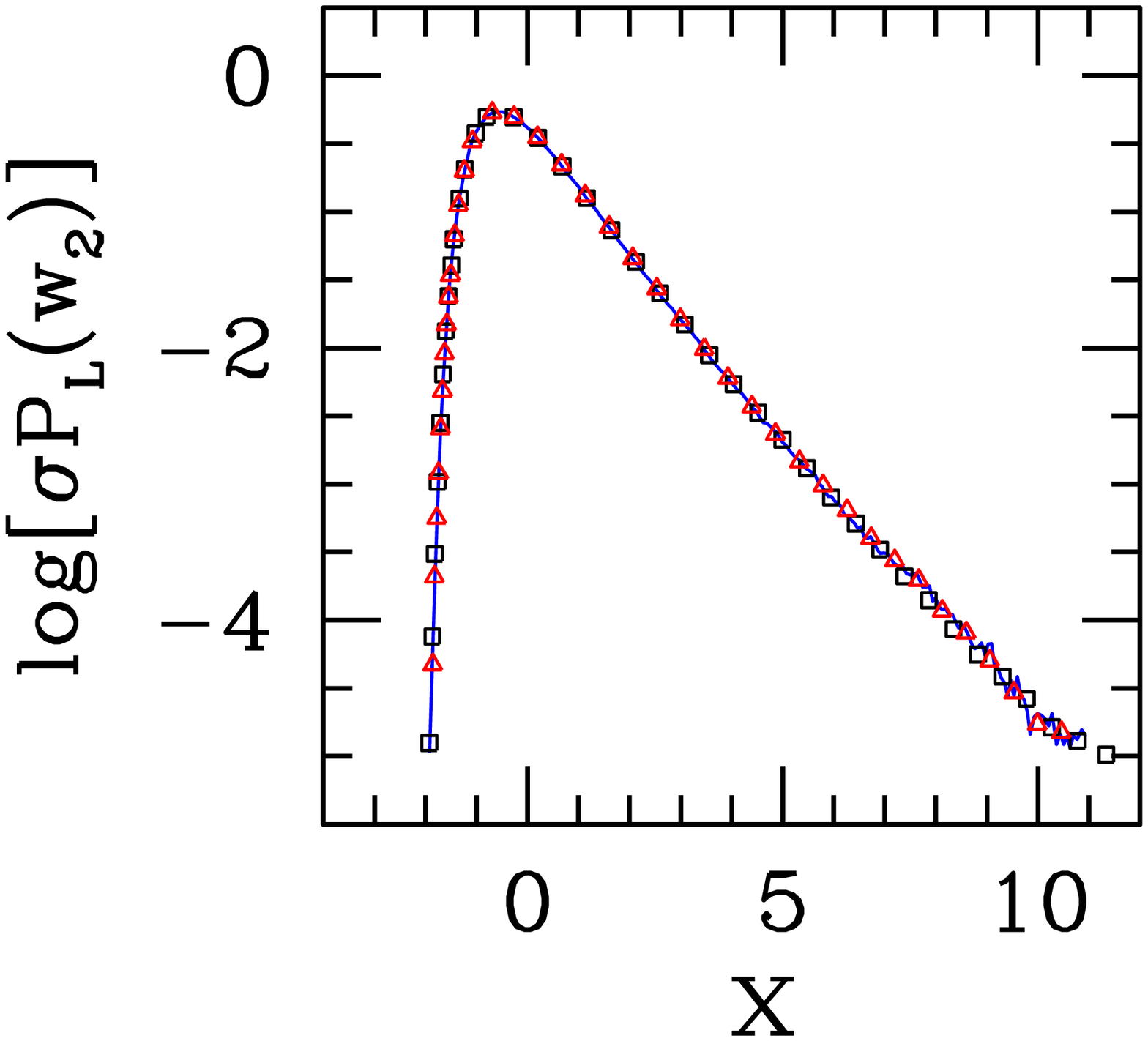}
\caption{\label{fig1} Scaled roughness distributions (according to Eq.
\ref{scaling1}) at the steady states of the models: RSOS in $L=256$
(solid line), etching in $L=256$ (squares) and RSOS2 in $L=128$
(triangles).}
\end{figure}     

Quantitative information on the shape of the KPZ distribution can be obtained
from the dimensionless ratios of the moments of the roughness distribution
$W_n\left( L\right) \equiv \int_0^\infty{ {\left( w_2 - \langle w_2\rangle
\right)}^n P_L\left( w_2\right) }$.
The ratios obtained with better accuracy involve the
lowest-order moments: the skewness
$S \equiv {{W_3}\over{{W_2}^{3/2}}}$
and the kurtosis
$Q \equiv {{W_4}\over{{W_2}^{2}}}-3$.
Although these quantities do not characterize the whole
distribution, they may be useful for the comparison with data from
experiments or other model systems if data collapse methods are
unreliable.

The finite-size estimates of $S$ and $Q$ for the distributions of
the RSOS and the etching models are shown in Figs. 2a and 2b.
Their asymptotic estimates are $S= 1.70\pm 0.02$ and $Q=5.4\pm 0.3$.
These large values show that the roughness distribution strongly
deviate from a Gaussian near the maximum. For $x<x_M$, where $x_M$ is the
abscissa of the maximum of the distribution in Fig. 1, the curve has an
approximately Gaussian shape centered at $x_M$. However, for $x>x_M$, it
significantly deviates from a Gaussian, showing a much slower
decay.

\begin{figure}[h]
\includegraphics{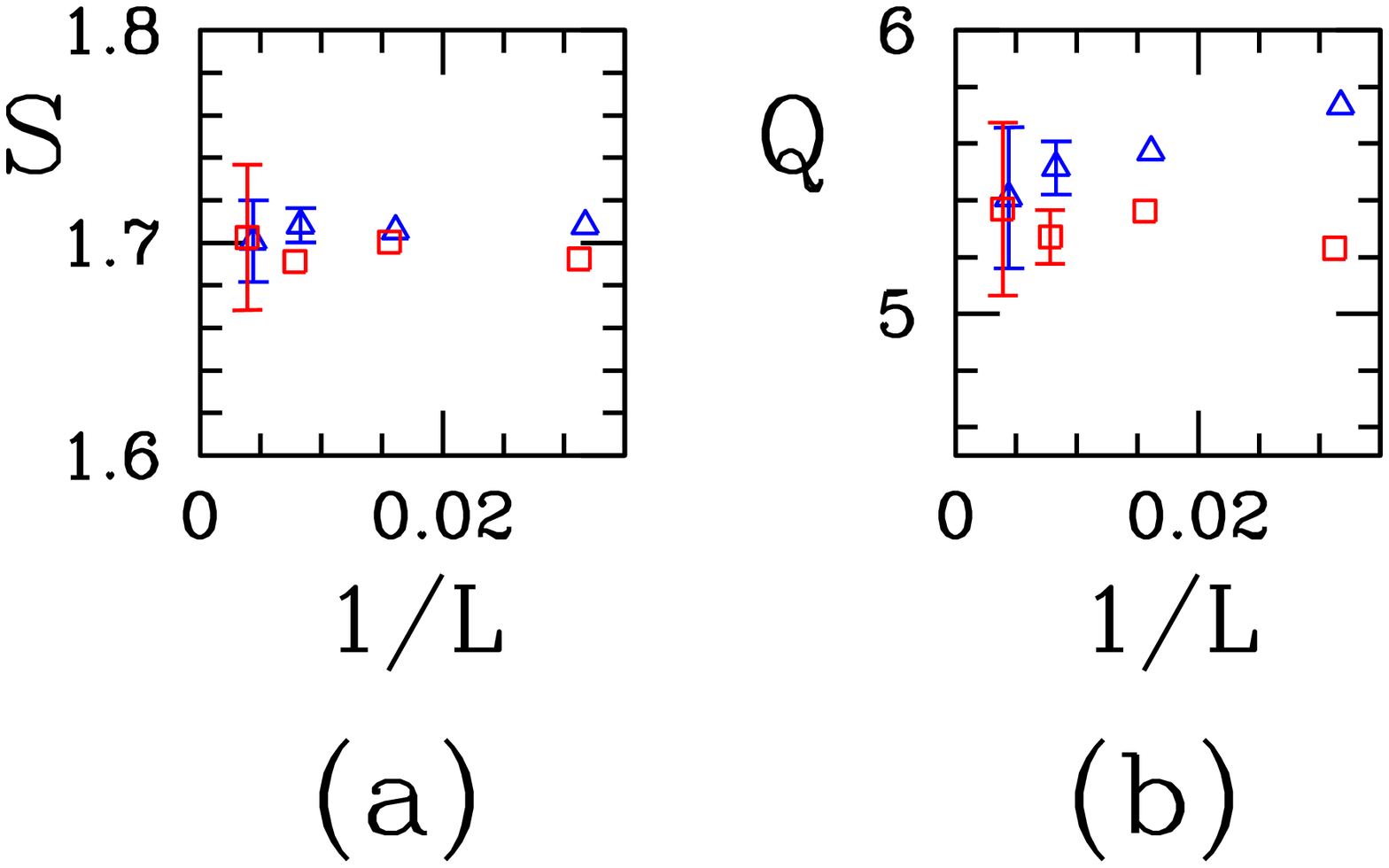}
\caption{\label{fig2}
(a) Skewness and (b) kurtosis of the interface width distributions at
the steady states of the RSOS (squares) and etching (triangle) models as a
function of inverse lattice length.}
\end{figure}     

The tail of the log-linear plot in Fig. 1 is
not a straight line, contrary to what is observed in the $1+1$-dimensional KPZ
model. This is certainly related to the non-Gaussian behavior of the KPZ
interface in $2+1$ dimensions~\cite{kpz2d,marinari}. It is reasonable to
assume that the scaling function decays as $\Psi\left( x\right) \sim x^\beta
\exp{\left( -Ax^\gamma \right)}$. Thus, for fixed $\beta$, the estimates
of the exponent $\gamma$ are given by
\begin{equation}
\gamma\left( x\right) = \frac{\ln{\left[
\frac{ \ln{\left( x^{-\beta} \Psi\left( x\right) \right)} }
{ \ln{\left( {\left( x-\Delta\right)}^{-\beta} \Psi\left( x-\Delta\right)
\right)} }
\right]}}
{\ln{\left[ x/\left( x-\Delta\right)\right]}} ,
\label{defgama}
\end{equation}
with constant $\Delta$.

In Fig. 3 we show the effective exponents obtained from the data of the
RSOS and the etching models in $L=256$ as a function of $1/x^2$, for
three values of $\beta$. The assumption of pure exponential
decay ($\beta =0$) leads to an asymptotic estimate $\gamma \approx 0.8$. With
$\beta = -0.2$, we obtain the smallest fluctuation of $\gamma\left( x\right)$
in the range of the variable $x$ analyzed here. Results for values of $\beta$
near $-0.2$ give $\gamma = 0.8\pm 0.1$. Although $\beta=-1$ suggests a simple
exponential decay ($\gamma =1$) for the etching model, it shows discrepancies
with the estimates for the RSOS model, reinforcing the proposal of the
stretched exponential form.

\begin{figure}[h]
\includegraphics{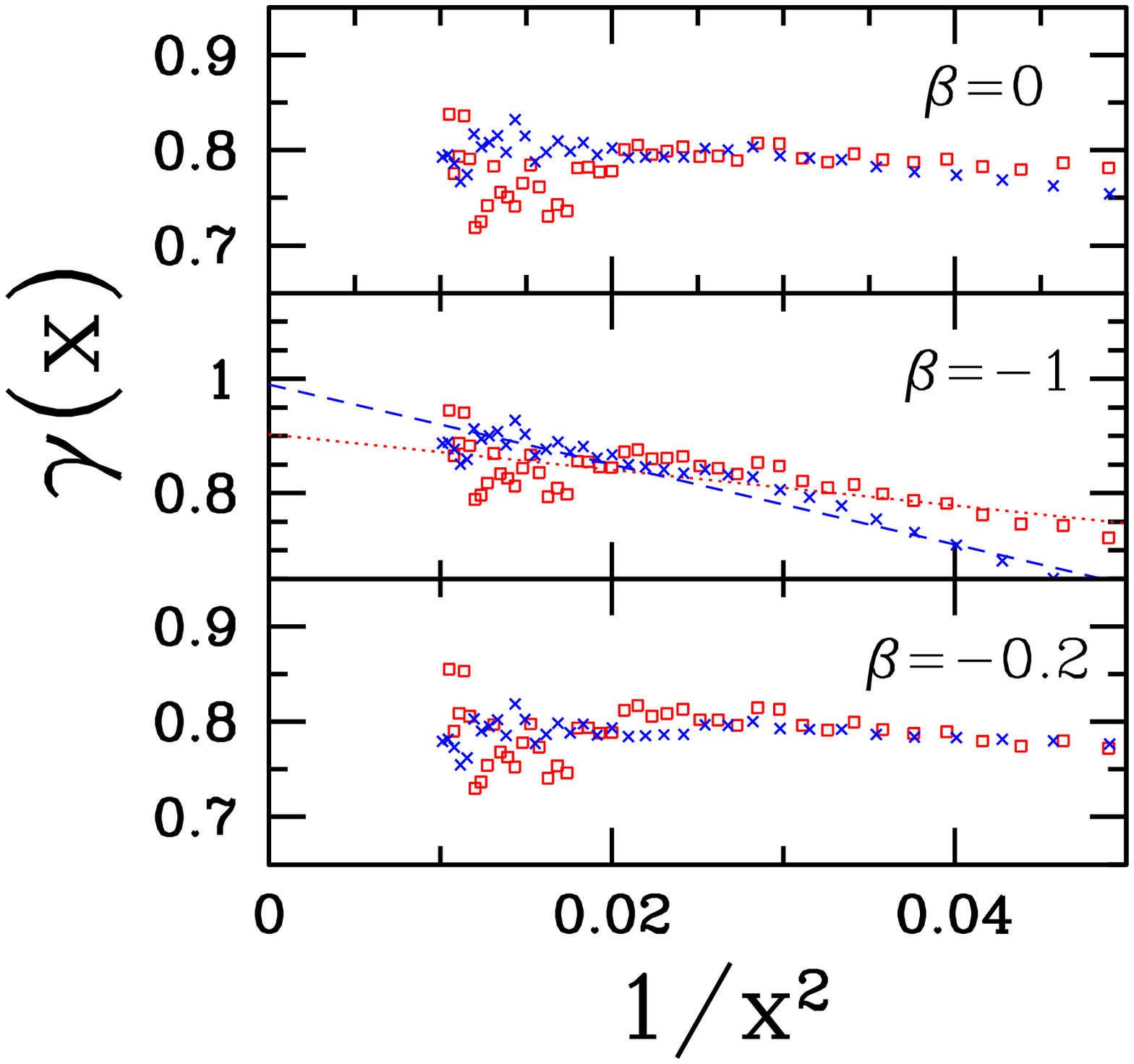}
\caption{\label{fig3}
Effective exponents $\gamma\left( x\right)$ of the exponential
decay of the scaling functions versus $1/x^2$, for three values of the
exponent $\beta$ and the models RSOS (squares) and etching (crosses). The
dotted (dashed) curve is a linear fit of the data of the RSOS (etching)
model.}
\end{figure}     

One question that could be raised here is the possibility of fitting the KPZ
distributions by those of $1/f^\alpha$ signals~\cite{antal}. However, the
plots of the distributions for several values of $\alpha$ show that this
procedure does not provide good fits of our data. Indeed, this can be
quantitatively explained by the above values of the skewness and the kurtosis,
which are much larger than the upper bounds for
the distributions of $1/f^\alpha$ signals~\cite{antal}:
$S=\sqrt{2}$ and $Q=3$, obtained in the limit $\alpha\to\infty$. Moreover, the
simple exponential decay of the $1/f^\alpha$ distributions for large $x$
disagrees with the $\exp{\left( -x^{0.8} \right)}$ decay of the KPZ models.

\section{Roughness distributions of VLDS models}

The CRSOS models considered here are extensions of the
original ones in which the incident particle executes a random walk
among neighboring columns until finding a position in which its
aggregation is allowed~\cite{reiscrsos}. In CRSOS1 and CRSOS2 models, the
conditions ${\Delta H}_{max}=1$ and ${\Delta H}_{max}=2$ between neighboring
columns are satisfied, respectively. Nearly ${10}^8$ configurations were used
to measured $w_2$ in each case.

Here, we will show distributions scaled according to Eq.
(\ref{scaling2}), as done in most previous works. This is also helpful for a
comparison with known results for the MH theory.

In Fig. 4a we plot $\log{\left[{\langle w_2\rangle}
P_L(w_2)\right]}$ versus $y\equiv \frac{w_2}{\left< w_2\right>}$
for the CRSOS1 and the CRSOS2 models in $d=1$, with
$L=256$. The scaling function for the MH theory, given in
Ref. \protect\cite{plischke}, is also shown. Its simple exponential decay
contrasts with the more rapid decay of the VLDS models.

The skewness and the kurtosis of the scaled VLDS distribution are obtained
along the same lines of the KPZ values: $S=1.22\pm 0.05$
and $Q=1.6\pm 0.2$. They contrast with the much larger values $S=1.988\dots$
and $Q=5.951\dots$ of the MH theory~\cite{plischke,antal}.
An approximately $\exp{\left( -y^2\right)}$ decay is suggested by a linear
fit of the data for the CRSOS1 model in the range $2\leq y\leq 5$, as shown
in Fig. 4b.

\begin{figure}[h]
\includegraphics{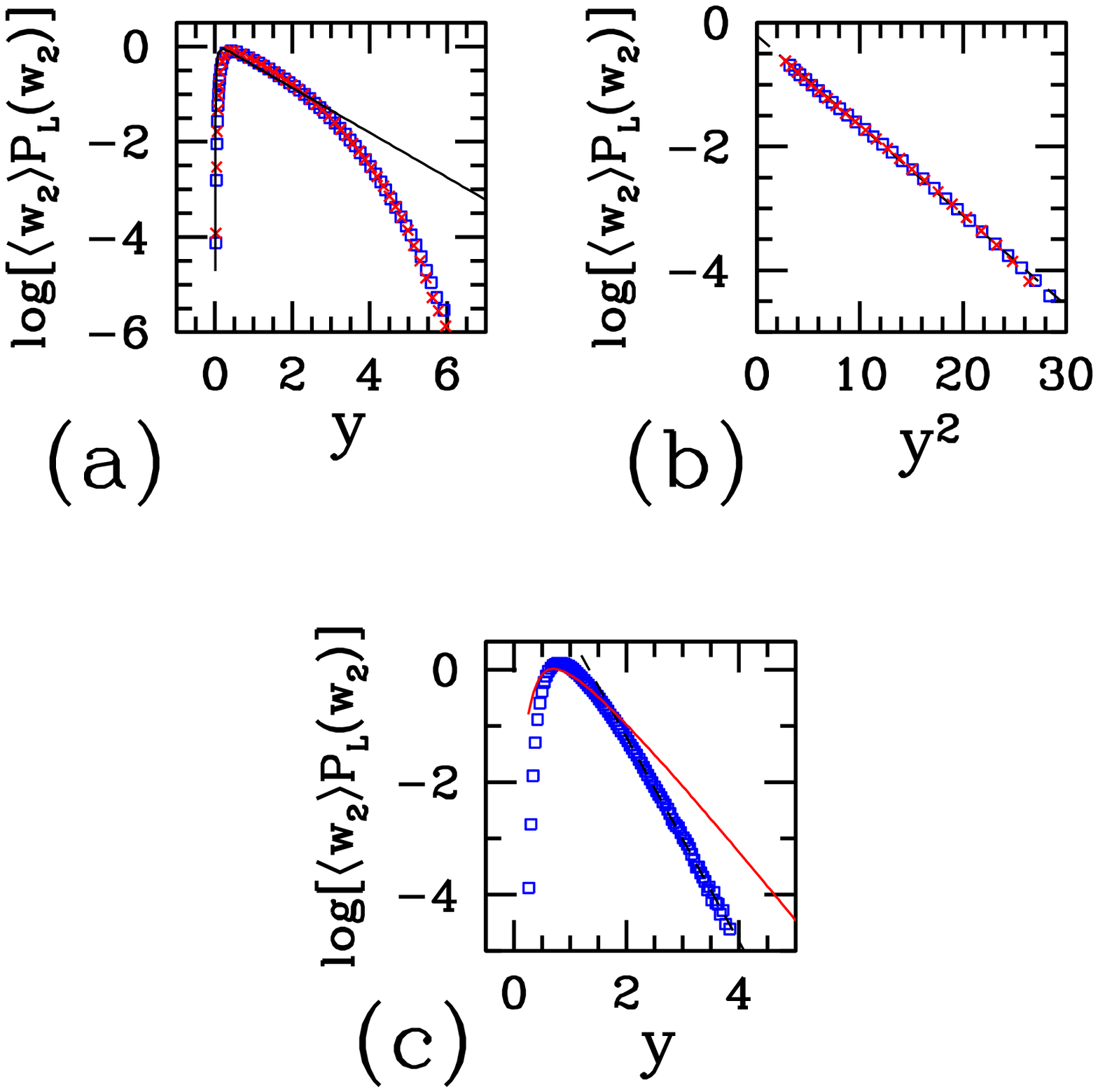}
\caption{\label{fig4}
(a) Scaled roughness distributions (according to Eq.
\ref{scaling2}) at the steady states of the CRSOS1 (squares) and CRSOS2
(crosses) models in $d=1$, with $L=256$. The solid curve is the
scaled distribution of the MH theory in $d=1$. (b) The same
distributions for CRSOS1 and CRSOS2 models as a function of $y^2$ and
a linear fit (dashed line) of the data for the CRSOS1 model.
(c) Distributions of the CRSOS1 model in $d=2$, with $L=64$, and a linear fit
of the data (dashed line) in the range $2<y<4$. The solid curve is the
distribution of the MH theory in $d=2$.}
\end{figure}     

Our results in $d=2$ are shown in Fig. 4c, where
scaled distributions of the CRSOS1 model ($L=64$) and the MH theory are
compared. The latter was obtained from the generating function given in Ref.
\protect\cite{racz} and the calculation of the residues of more than $20$
poles closest to the origin. Both distributions have a simple exponential
decay for large $y$, which is justified in the VLDS case by the linear fit
shown in Fig. 5 ($2\leq x<4$). However, the VLDS distribution is much narrower
than that of the MH theory: the maximum of the VLDS scaling function is $\Phi
\approx 1.3$, while the maximum of the MH one is $\Phi\approx 1$ (see, e. g.
Ref. \protect\cite{racz}), and the respective decays are, approximately,
$\exp{\left( -1.8y\right)}$ and $\exp{\left( -1.3y\right)}$. For the VLDS
distribution, the skewness and kurtosis are $S\approx 1.1$ and $Q\approx 1.8$,
while the values for the MH theory are $S\approx 1.30$ and $Q\approx 2.63$.

We also tried to fit the scaled distributions of the CRSOS1 model, in $d=1$
and $d=2$, with $1/f^\alpha$-noise distributions~\cite{antal}, with some value
of $\alpha$. In $d=1$, no good fit was already
expected due to the form of the decay of the scaling function discussed
above. In $d=2$, there is no simultaneous agreement of the values of $S$ and
$Q$ of those distributions and our estimates, for any $\alpha$, indicating
that this type of fit is also impossible. In Ref. \protect\cite{racz}, a fit
with $\alpha=3$ was suggested for the distribution of a growth model with
Arrhenius dynamics, but this was probably a consequence of the lower accuracy
of the data for that model.

\section{Conclusion}

Roughness distributions were calculated for several discrete models in
the KPZ and the VLDS class, in one- and two-dimensional substrates, confirming
the scaling relations (\ref{scaling2}) and
(\ref{scaling1}). The scaling functions are different
from their counterparts of the linear theories, EW (second order) and MH
(fourth order). In the $2+1$-dimensional KPZ class, large values of the
skewness $S$ and kurtosis $Q$ of the scaling functions are related to the
high asymmetry near the maximum. For large $x$, the scaling functions decay
approximately as $\exp{\left( -x^{0.8}\right)}$, in contrast to the simple
exponential decay of the linear theories and of $1+1$-dimensional KPZ. In the
$1+1$-dimensional VLDS class, $S$ and $Q$ are smaller than those of the MH
theory and a decay approximately as $\exp{\left( -x^2\right)}$ is obtained.
For the $2+1$-dimensional VLDS class, the simple exponential decay of the
scaled distribution is quantitatively different from that of the MH theory.

Another interesting point of our study is that none of the above
distributions are fitted by those of
$1/f^\alpha$ noise, for any $\alpha$. It contrasts with the
distributions for model and experimental interfaces in the depinning
thresholds analyzed in recent works~\cite{moulinet,dequeiroz,rosso}, which
indicates strong effects of the non-Gaussian nature of the KPZ (in $d=2$) and
VLDS interfaces. On the other hand, it suggests the study of other
models with depinning transitions, particularly those with
relatively simple dynamics whose growth phases are in the KPZ class~\cite{bjp},
for a comparison with $1/f^\alpha$ noise distributions.

\begin{acknowledgments}

The author thanks Prof. Zoltan R\'acz and Prof. Jorge de S\'a Martins for
helpful discussions and suggestions.

This work was partially supported by CNPq and FAPERJ (Brazilian agencies).

\end{acknowledgments}


\end{document}